\documentclass[floatfix,showpacs,amsmath,amssymb,letterpaper,groupaddresses,superscriptaddress]{article}
\usepackage{latexsym}
\usepackage{epsfig}
\usepackage{mathtools}
\usepackage{amsmath}
\usepackage{amssymb}
\usepackage{graphicx}
\usepackage{times}
\usepackage[section]{placeins}
\usepackage{caption}
\usepackage{subcaption}
\usepackage{bm}
\usepackage{color}

\newtheorem{definition}{Definition}

\def\a{\alpha}
\def\b{\beta}

\def\be{\begin{equation}}
\def\ee{\end{equation}}
\def\ba{\begin{eqnarray}}
\def\ea{\end{eqnarray}}
\def\la{\langle}
\def\ra{\rangle}
\def\a{\alpha}

\def\b{\beta}

\def\h{\hskip 1cm}

\def\lo{\longrightarrow}

\begin{document}

\vspace{4cm}
\begin{center}{\Large \bf  Entangled States as Robust and Re-usable Carriers of Information}\\
\vspace{2cm}

Shima Emamipanah $^b$, Marzieh  Asoudeh$^b$\footnote{Corresponding Author} and Vahid Karimipour.$^a$\\
\vspace{1cm} $^a$ Department of Physics, Sharif University of Technology, P.O. Box 11155-9161, Tehran, Iran.\\
$^b$Department of Physics, Azad University, Northern Branch,  Tehran, Iran .\\

\end{center}

\begin{abstract}
  Entangled states can be used as secure carriers
  of information much in the same way as carriers are used in  classical communications. In such protocols, quantum states are uploaded to  the carrier at one end and are downloaded from it in safe form at the other end, leaving the carrier intact and ready for reuse. Furthermore, protocols have been designed for performing quantum state sharing in this way. In this work, we study the robustness of these protocols against  two of the most common sources of noise, namely de-phasing and depolarization  and show that multiple uses of these carriers do not lead to accumulative errors, rather the error remains constant and under control.
\end{abstract}

\section{Introduction}\label{intro}
Conventionally in quantum information processes \cite{tele, dense, mbqc, blind} entanglement is used as a resource which is consumed at the end of a process and has to be renewed for a second use. This is the case for teleportation, measurement-based quantum computation  and many other protocols. For example in quantum secret sharing schemes \cite{buz, Karlsson, chin1, chin2, chin3, chin4}, which is the subject of interest in the present work, the highly non-classical correlation in the shared entangled state allows the legitimate parties to establish a random shared key between themselves. Of course there are also cryptographic protocols which are sequential and do not use entanglement at all \cite{BB84, Ek, Dagmar, Sch, ZhangLiMan, Tavak, KA}.\\

The idea of using an entangled state between two remote points, as a reusable carrier of information, first came up in \cite{ZhangBasic} and then extended to quantum secret sharing  in \cite{BK} by one of the authors. This idea is a natural generalization of the idea behind today's classical communication networks, in the sense that an entangled state between two or more  points acts as a secure carrier of information between these points.  The sender entangles (uploads) a state to the carrier which is disentangled (downloaded) by the receiver at the other end. The additional feature, due to the quantum properties is that during  transmission, the state is hidden from potential adversaries.  The hiding effect is a direct result of entanglement of the message state with the carrier state, by which the message state is in a highly mixed state and carrying no information at all by itself.  Only in the sender and receiver ends, where the message is uploaded (entangled to the carrier) or downloaded (disentangled from the carrier), the identity of the message is revealed.  At the end of each round, the carrier returns to its initial state and is ready for a second round of use. We stress that an entangled state is not used here in the usual sense, i.e. to securely establish a shared random key between the parties\cite{Gisin}, but to send a deterministic message from one end to the other. In  case of \cite{ZhangBasic} the carrier is a simple Bell state between two parties, while in  case of \cite{BK}, the carrier is a three party entangled state and alternates between two specific forms in consecutive rounds. In this protocol Alice sends a classical or quantum message to Bob and Charlie who can retrieve it only by their collaboration. Note that by classical message we mean classical bits which have been encoded into computational states $\{|0\ra , |1\ra \}$ and by quantum message we mean an unknown quantum state $|\phi\ra=\a|0\ra+\beta|1\ra.$\\

We should stress that entanglement in this sense, that is, for sending deterministic messages, has not been reported previously before the  two references \cite{ZhangBasic} and \cite{BK}. Of course the price of this higher level protocol is the bigger resource that the players have to use, i.e. the multi-party entangled states in the transition stage. These schemes may seem impractical with present technologies, however with the intensive research on quantum networks, they may become potential candidates for the quantum communication methods  of tomorrow. Therefore it is important to study these schemes from different points of view. In particular it is important to investigate how resilient these protocols are against noise of the environment. Of particular importance are the de-phasing and depolarizing  noise  which we study in this paper.  We will examine in detail the scheme of \cite{BK} for Quantum State Sharing (QSS) and determine how much error the de-phasing and depolarizing noise incur on the transmitted data and states.   It is important to determine how much error the noise of environment will incur on the fidelity between the sent and the received q(bits). This will allow the legitimate players to distinguish the noise resulting from the action of an adversary from the environmental noise and help them to reveal the presence of an adversary. Note that our emphasis is on the concept of entangled state as a reusable carrier of information and not on the security of such a protocol against various types of attacks. 
For discussions on two kinds of attacks on this protocol see \cite{attack1, attack2} and \cite{my1, my2}. For generalization of this protocol to polynomial codes and $(k,n)$ threshold schemes, see \cite{marvian} where  further discussions on the performance of the protocol under various  outsider attacks and cheating of insiders, shas been made.   \\

An important question is whether or not a noisy carrier, specially when it is continually under the uploading,  downloading and Hadamard actions of the players, will keep its shape or will be disrupted after a few rounds and will be completely useless.  We will show that this is not the case and the carrier will  keep its form and is only mixed by a well-known and controlled state.  Here we will study the case where the carrier has been contaminated by noise before the operations start and the time scale of operations in each round are short enough that the additional noise during the operations, either on the carrier or on the encoded states are negligible.\\                                                                                                    	

The structure of the paper is as follows: In section (\ref{rev}), we briefly review the protocols of  \cite{BK} in a self-contained way. The main point of the scheme of \cite{BK} is that the three parties involved in the state sharing scheme, have to use two kinds of carriers in alternating rounds and these two carriers are turned into each other by their local action at the end of each round.  In sections (\ref{n2}) and section (\ref{n3}) we study the effect of two important types of noise, namely de-phasing and depolarizing noise on this protocol. Although we focus on the three-party case, the idea can be easily generalized to schemes with more than three players. We will briefly discuss this in section (\ref{n}), where we also point out the possibility of error correction schemes for overcoming the effect of aforementioned noise.  We end the paper with conclusion and outlook. \\

\section{Entangled states as reusable carriers of information}\label{rev}

Consider three parties, Alice (A), Bob (B) and Charlie (C). The goal of Alice is to send a state $|\psi\ra=a|0\ra+b|1\ra$ to Bob and Charlie so that they can retrieve it only if they collaborate with each other. We assume that Bob and Charlie are at one location and can affect two qubit gates on the state they receive. In the meantime Alice wants to keep this state from any possible adversaries by entangling it with a carrier which is shared between her and the other two parties. In the end the carrier should be such that it can be used again for sending other qubits.  Before proceeding, it is better that we collect all our notations and conventions and some basic facts and formulas which are easily proved by direct calculations. \\

\subsection{Notations and conventions}
We use the subscripts $A, B, $ and $C$ for the qubits of carrier possessed by Alice, Bob and Charlie and the subscripts $1$ and $2$ for the two qubits which are being communicated. The qubit 1 is always sent to Bob and qubit 2 is always sent to Charlie. In equations and for the sake of simplicity, we write these subscripts only on the left hand side. The same subscripts are to be understood for the states on the right hand side. \\

\noindent Let $q$ be a bit $\in \{0,1\}$. $\overline{q}$ is meant to denote the negation of $q$, i.e. $\overline{0}=1, \overline{1}=0$. We define the  state
\be
|{\bf 0}_2\ra=\frac{1}{\sqrt{2}}(|00\ra+|11\ra),\h |{\bf 1}_2\ra=\frac{1}{\sqrt{2}}(|01\ra+|10\ra),\ee
to represent respectively the two qubit states of even and odd parity. In close form we write
\be |{\bf q}\ra=\frac{1}{\sqrt{2}}(|0q\ra+|1\overline{q}\ra).\ee

{\bf Remark:}{\it{
Hereafter and most of the time, we write all states without normalization and assume that they are understood to be properly normalized. Thus, a state like $|000\ra+|111\ra$ is to be understood as
$\frac{1}{\sqrt{2}}(|000\ra+|111\ra)$.}}\\

The even and odd parity states can be extended to more than two parties. For three qubits we have
\ba\label{33}
|{\bf 0}_3\ra&=&|0\ra|{\bf 0}_2\ra+|1\ra|{\bf 1}_2\ra,\cr
|{\bf 1}_3\ra&=&|0\ra|{\bf 1}_2\ra+|1\ra|{\bf 0}_2\ra.
\ea
This can be extended to $n$ parties in the same way by induction on $n$, although we do not need this here. The subscripts A, B and C are used to denote qubits in possession of the three players conventionally called Alice, Bob and Charlie. The subscripts 1 and 2 are used to denote the qubits which are being transmitted, one to B and the other to C. We write these subscripts only on the left hand side of equations, with the understanding that they also apply to the states on the right hand side, that is $$|\psi\ra_{A,B}=|00\ra+|11\ra
$$ is meant to denote $$|\psi\ra_{A,B}=|00\ra_{A,B}+|11\ra_{A,B}.$$
When from the context it is  clear what types of parity state we are dealing with, we refrain from writing the subscript $2$ or $3$ on ${\bf 0}$ or ${\bf 1}$. For example it is clear that a state $|{\bf 0}\ra_{A,B,C}$ is a $|{\bf 0}_3\ra$ state and a state like  $|{\bf 0}\ra_{1,2}$ is a $|{\bf 0}_2\ra$ state. \\

\begin{figure}[!ht]
	\centering
	\includegraphics[width=12.0cm,height=10cm,angle=0]{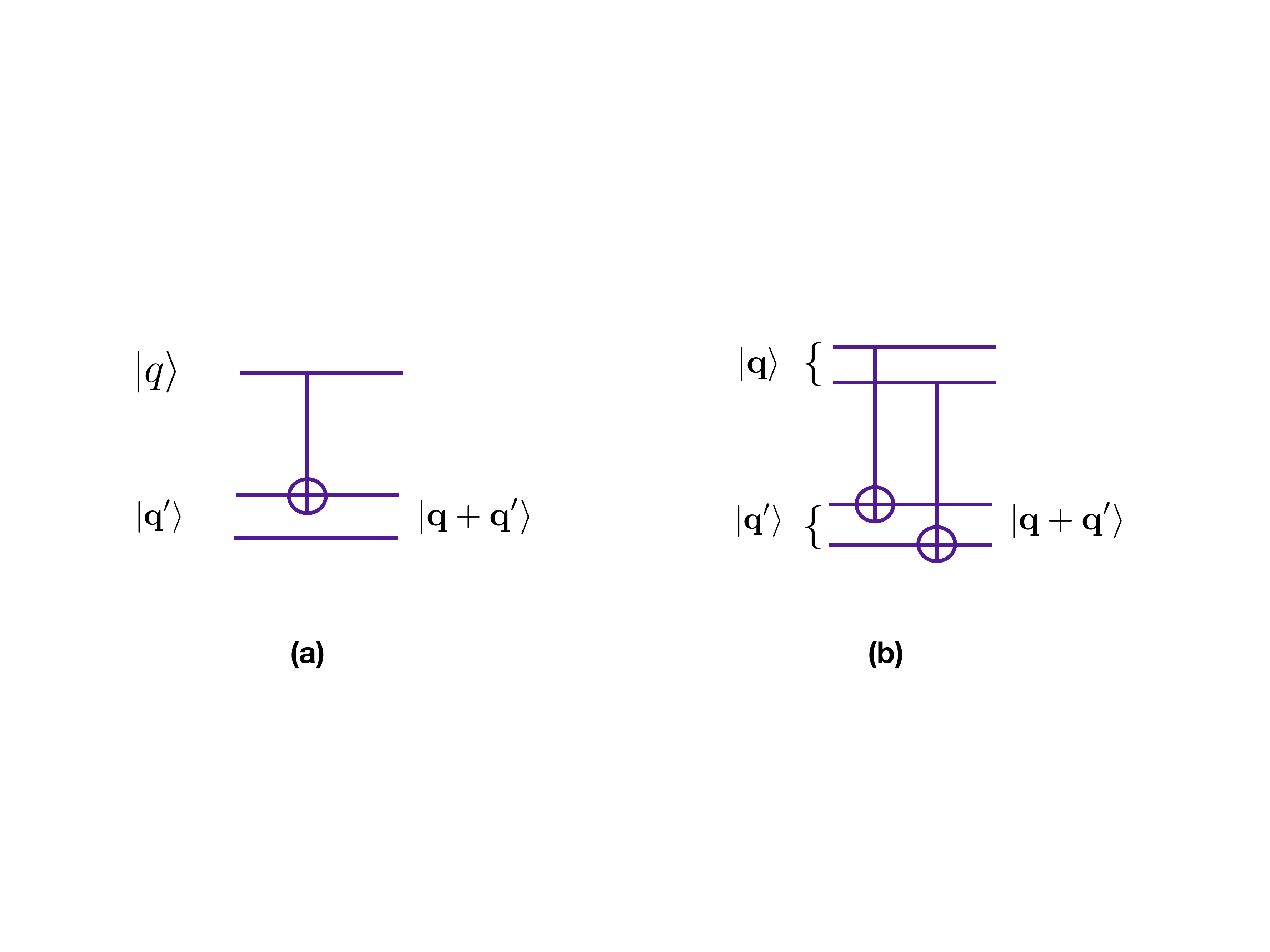}\vspace{-2.5cm}
	\caption{ The CNOT operators on encoded states, (a): equation (\ref{CXX}), (b): equation (\ref{CYY}).
	}\label{CNOT}
\end{figure}

\noindent A CNOT operation with control bit $X$ (where $X\in \{A, B, C\}$) and target bit $1$ is denoted by $C_{X1}$. Simple calculation shows that on this encoded states \cite{BK}, we have the following CNOT operations, Fig. (\ref{CNOT})
\be\label{CXX}
\left(C_{X,1} \ {\rm or }\ C_{X,2}\right)|q\ra_X|{\bf q'}\ra_{1,2}=|q\ra_X|{\bf {q+q'}}\ra_{1,2},
\ee
and
\be\label{CYY}
C_{X,1}C_{Y,2}|{\bf q}\ra_{X,Y}|{\bf q'}\ra_{1,2}=|{\bf q}\ra_{X,Y}|{\bf q+q'}\ra_{12}.
\ee
Note the difference: In Eq. (\ref{CXX}) only one of the CNOTs should be applied, while in Eq. (\ref{CYY}) both CNOT's should be applied.
 \begin{figure}[!ht]
	\centering
\includegraphics[width=12.0cm,height=10cm,angle=0]{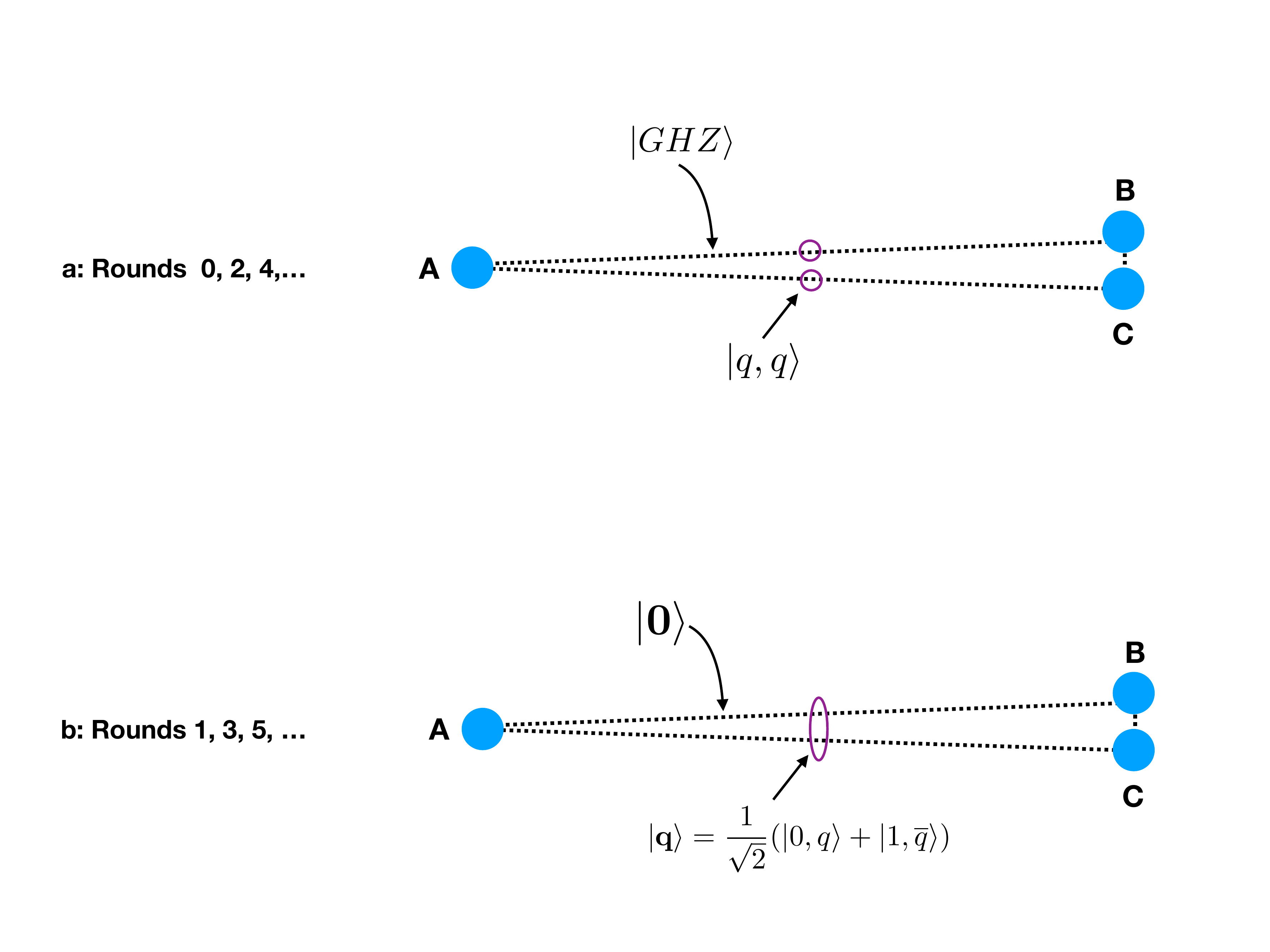}
	\caption{ In even rounds $0, 2, 4, \cdots$, the carrier is $|GHZ\ra$ and $q$ is encoded into a product state $|q,q\ra$. In odd rounds $1,3,5,\cdots $ the carrier is $H^{\otimes 3}|GHZ\ra=|{\bf 0}\ra$ and $q$ is encoded into an entangled state $|{\bf q}\ra_{1,2}.$ In the even rounds, where Bob and Charlie can retrieve the bit independently,  Alice can send redundant bits which carry no information.
	}\label{carriers}
\end{figure}
Finally we numerate the rounds as $0, 1, 2, \cdots$ The superscript  $^{(0)},^{(1)}, ^{(2)},\cdots $ are used to denote the form of the carrier at the start of rounds $0, \ 1. , \ 2, \cdots $.\\

\subsection{The QSS scheme in the absence of noise}
The three party secret sharing scheme \cite{BK}, which can easily be extended to $N-$party scheme runs as follows
 Alice, Bob and Charlie (A,B and C) share a GHZ state
\be\label{GHZ}
|GHZ\ra_{_{A,B,C}}=\frac{1}{\sqrt{2}}(|000\ra+|111\ra),
\ee
which upon the joint Hadamard action of all three parties $H_A\otimes H_B\otimes H_C$ turns into another carrier

\be
|{\bf 0}\ra_{_{A,B,C}}=\frac{1}{2}\left(|000\ra+|110\ra+|101\ra+|011\ra\right),
\ee
Since $H^2=I$, we have

\be
H_A\otimes H_B\otimes H_C|{\bf 0}\ra_{_{A,B,C}}=|GHZ\ra_{_{A,B,C}},\h H_A\otimes H_B\otimes H_C|GHZ\ra_{_{A,B,C}}=|{\bf 0}\ra_{_{A,B,C}}.
\ee
Therefore  the two types of carriers alternate in even and odd rounds (starting from round 0 with GHZ). Consider now the even rounds 0, 2, 4, $\cdots$, where the carrier is $|GHZ\ra_{_{A,B,C}}$ and Alice encodes $|q\ra$ to $|qq\ra_{1,2}$ and entangles this state to the carrier by the operation $C_{A,1}\otimes C_{A,2}$. The total state becomes
\be\label{Psieven}
|\Psi_{even}\ra_{_{A,B,C,1,2}}=|000\ra|qq\ra+|111\ra|\overline{q}\overline{q}\ra.
\ee
At the other end,  Bob and Charlie disentangle the state $|qq\ra$ by operations
$C_{B,1}$ and $C_{C,2}$, returning the carrier to the $|GHZ\ra$ state,
and read the message bit $q$ independently. In these rounds, the sequence of operations can be collected
in the form

\be
\Omega^{even}=C_{_{C,2}}C_{_{B,1}}C_{_{A,2}}C_{_{A,1}}.
\ee
At the end of even rounds, all three parties act by their Hadamard operation on their share and turn the carrier to the state $|{\bf 0}\ra_{_{A,B,C}}$. In the odd rounds 1, 3, 5, $\cdots $, where the carrier is $|{\bf 0}\ra_{_{A,B,C}}=|0\ra_{_A}|{\bf 0}\ra_{_{B,C}}+|1\ra_{_A}|{\bf 1}\ra_{_{B,C}}$, Alice encodes the state $|q\ra$ into $|{\bf q}\ra_{1,2}$ and entangles this state to the carrier by her operation $C_{_{A,1}}$ on only one of the qubits (see Eq.(\ref{CXX})). This changes the total state to

\be
|\Psi_{odd}\ra_{_{A,B,C,1,2}}=|0\ra_A|{\bf 0}\ra_{_{B,C}}|{\bf q}\ra_{1,2}+|1\ra_A|{\bf 1}\ra_{_{B,C}}|{\overline{\bf q}}\ra_{1,2}.
\ee

\noindent Again  the state in travel is a mixture of $|{\bf q}\ra$ and $|{\bf \overline{q}}\ra$ and is hidden from any adversary. At the destination, Bob and Charlie perform the operation $C_{B,1}C_{C,2}$ to disentangle the state from the carrier (download it) (see (Eq. \ref{CYY})). In these rounds, the sequence of operations can be collected
in the form

\be
\Omega^{odd}=C_{_{C,2}}C_{_{B,1}}C_{_{A,1}}.
\ee

But this time the state can be decoded only if Charlie and Bob collaborate with each other (i.e. Acting by $C_{1,2}$ on the received state).  It has already been shown in \cite{BK} that Eve cannot entangle herself with the carrier state, since the joint Hadamard operation effectively throws her out by revealing her presence to the legitimate parties. \\

{\bf Remark:}{\it{
Any basis state $|q\ra$ can be sent by Alice and retrieved by the receivers by encoding and decoding it differently  in even and in odd rounds. Therefore by linearity of the process, this is also true for  any linear combination of these basis states, i.e. for any unknown quantum state $|\psi\ra_{1,2}=\a|0\ra+\b|1\ra$ which is encoded to  $|\overline{\psi}\ra_{1,2}=\a|0,0\ra+\b|1,1\ra$ in even rounds and to $|\overline{\psi}\ra_{1,2}=\a|{\bf 0}\ra+\b|{\bf 1}\ra$ in odd rounds. Note that the even rounds is not used for sharing data between Bob and Charlie and in fact  Alice can spare these rounds and send sensitive data only in odd rounds where the receivers should collaborate to retrieve the data. The answer to the question why the parties do not always use the odd-type carrier is that the Hadamard operations are necessary for the security of the protocol \cite{BK} and the action of these Hadamard operations inevitably changes the carrier in alternate rounds. }}\\

\begin{definition}\label{defff}
	 By the complete channel $\Phi$ between the sender Alice and one of the receivers, say Bob, we mean the channel which consists of encoding an initial state $|\psi\ra=\a|0\ra+\b|1\ra$ to either $|\psi_{even}\ra=\a|00\ra+\b|11\ra$ or $|\psi_{odd}\ra\a|{\bf 0}_2\ra+\b|{\bf 1}_2\ra$, sending it through the quantum channel described above and then decoding to the initial state $|\psi\ra$ by the collaboration of Bob and Charlie. In the absence of noise, this channel is an identity qubit channel, i.e. $\Phi(\rho)=\rho.$\\
\end{definition}

We are now ready to study the robustness of this scheme first under De-phasing and then under global depolarizing noise. In other words, we want to find out what happens to the complete channel defined above, when the carriers are noisy and are not pure states any more.  \\

\section{Effect of de-phasing noise}\label{n2}
A ubiquitous form of noise is De-phasing which maps the carrier state $|GHZ\ra$ into the following
\be\label{rho-0}
\rho^{(even)}_{_{A,B,C}}=(1-2p)|GHZ\ra\la GHZ|+p|000\ra\la 000|+p|111\ra\la 111|.
\ee
Such a state is the result of random phase kicks on the three qubits. In fact  random operation of a product  operator   $$U_{_{A,B,C}}=e^{i\theta_1\sigma_{z}}\otimes e^{i\theta_2\sigma_{z}}\otimes e^{i\theta_3\sigma_{z}}$$
with probability $P(\theta_1,\theta_2,\theta_3)$ produces a state in the form
\be
\int dU\ U|GHZ\ra\la GHZ|U^{\dagger},
\ee
where $dU=P(\theta_1,\theta_2,\theta_3)d\theta_1d\theta_2d\theta_3$. A simple calculation shows that this produces the state in (\ref{rho-0}) with the following parameter $p$

\be
p=\frac{1}{2}\big[1-\int d\theta_1d\theta_2d\theta_3 e^{2i(\theta_1+\theta_2+\theta_3)}P(\theta_1,\theta_2,\theta_3)\big].
\ee
Here we have used the plausible assumption that the random kicks are acting on qubits as single gates, although their probability need not be uncorrelated. Moreover we have assumed that the probability distribution function is an even function of its arguments, that is, the phase kicks are symmetric around zero. \\

As we will see in the sequel, it is convenient if we define another GHZ state in the form

\be\label{GHZ'}
|GHZ'\ra_{_{A,B,C}}=\frac{1}{\sqrt{2}}(|000\ra-|111\ra).
\ee
We can then rewrite the noisy state $\rho^{(0)}$ in the form

\be\label{new-rho-0}
\rho^{(even)}_{_{A,B,C}}=(1-p)|GHZ\ra\la GHZ|+p|GHZ'\ra\la GHZ'|.
\ee

The  main concern  is that, with the operations of the three parties (i.e. the uploading, downloading and Hadamard operations), the shape of the carrier may continually changes in each round and  becomes useless for transmission of states after a few rounds.  However as we will show in the sequel, this is not the case. To see this we first note from (\ref{Psieven}) and (\ref{new-rho-0}) that in even rounds both carriers $|GHZ\ra$ and $|GHZ'\ra$ are equally effective in delivering a state $|qq\ra$ from A to B and C in safe form. Thus, in the even rounds de-phasing noise has no detrimental effect on the scheme. At the end of these rounds, the state is acted upon by the Hadamard operations $H_A\otimes H_B\otimes H_C$ and turns into

\be\label{oddd}
\rho^{(odd)}_{_{A,B,C}}=(1-p)|{\bf 0}\ra\la {\bf 0}|+p|{\bf 1}\ra\la {\bf 1}|.
\ee

To analyze the performance of this carrier, we need only study the carrier $|{\bf 1}\ra_{ABC}$ and then combine the results, since we already know that the carrier $|{\bf 0}\ra_{ABC}$ delivers the state $|{\bf q}\ra$ in safe form. In odd rounds, the operations of the players is given by $\Omega^{odd}=C_{_{C,2}}C_{_{B,1}}C_{_{A,1}}$. This operation is shown in circuit diagram of Fig. (\ref{CircuitOdd}). We have to affect this operator on
\be\label{abc}|{\bf 1}\ra_{_{A,B,C}}|{\bf q}\ra_{_{1,2}}\equiv X_AX_BX_C|{\bf 0}\ra_{A,B,C}|{
\bf q}\ra_{1,2}.\ee
 To affect this, we note from Fig. (\ref{CircuitOdd}) that $$\Omega^{odd}X_{_A}X_{_B}X_{_C}=X_2X_{_A}X_{_B}X_{_C}\Omega^{odd}$$ which in combination with (\ref{abc}) and the fact that $X_2|{\bf q}\ra=|\overline{\bf q}\ra$,
 shows that in odd rounds

\be
|{\bf 1}\ra_{_{A,B,C}}|{\bf q}\ra_{_{1,2}}\lo |{\bf 1}\ra_{_{A,B,C}}|\overline{{\bf q}}\ra_{_{1,2}}.
\ee

\begin{figure}[!ht]
	\centering	\includegraphics[width=12.0cm,height=10cm,angle=0]{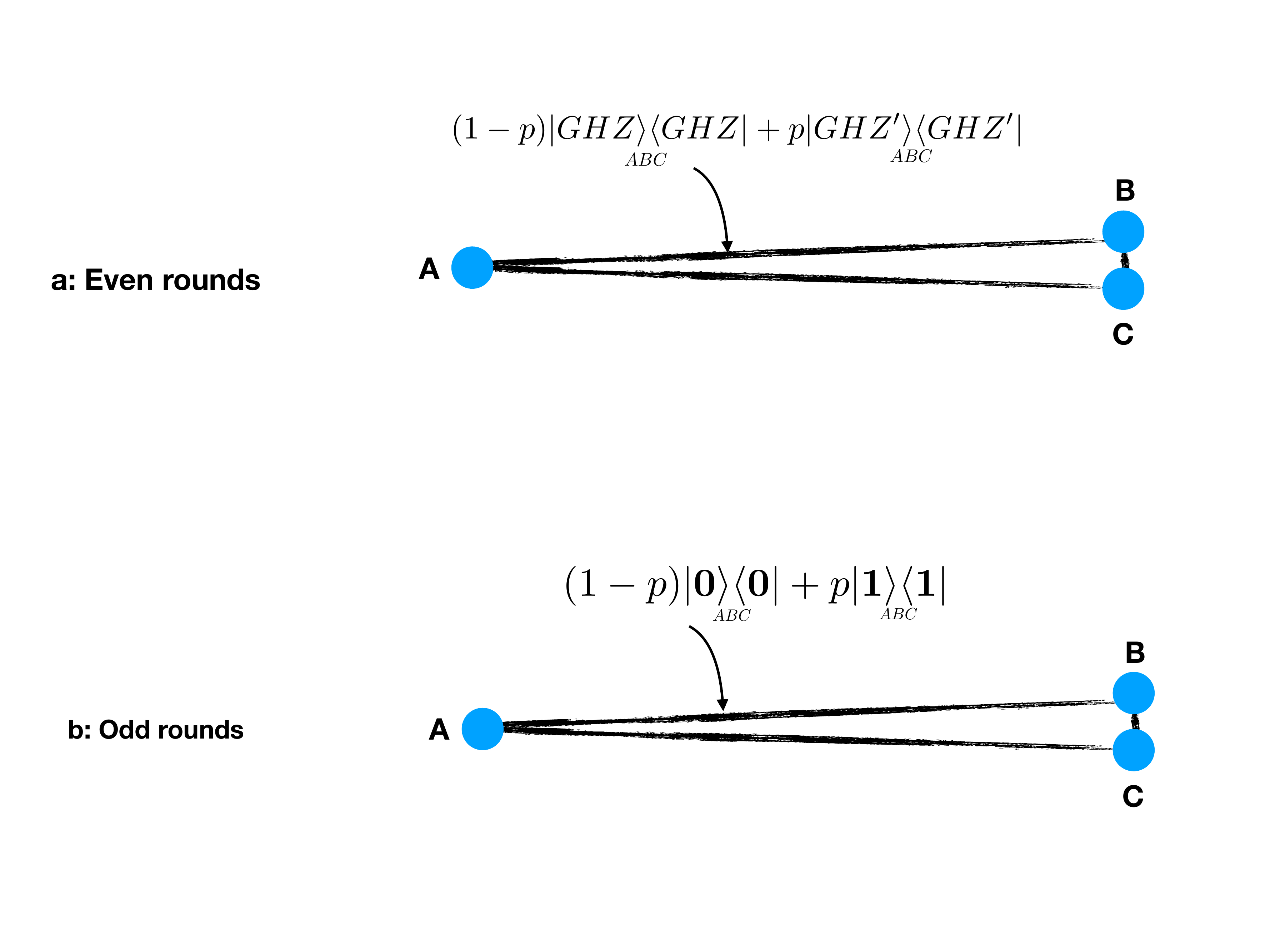}
	\caption{ In even rounds, where $q$ is encoded into a product state $|qq\ra_{1,2}$ the carrier $|GHZ\ra_{_{A,B,C}}$ is contaminated by $|GHZ'\ra_{_{A,B,C}}$ which acts as good as $|GHZ\ra_{_{A,B,C}}$. In odd rounds where $q$ is encoded into an entangled state $|{\bf q}\ra$,  the carrier $|{\bf 0}\ra_{_{A,B,C}}$ is contaminated by $|{\bf 1}\ra_{_{A,B,C}}$ which flips the state to $|\overline{{\bf q}}\ra$. The two kinds of carriers turn into each other after each round.
	}\label{nsf}
\end{figure}

This means that
in even rounds the message state being encoded as $|qq\ra$ is delivered in safe form and in odd rounds the message state encoded as $|{\bf q}\ra$ is delivered with probability $(1-p)$ in safe form as $|{\bf q}\ra$ and with  error probability equal to  $p$ as $|\overline{{\bf q}}\ra$. The important point is that the sequence of operations do not change the carriers anymore except the Hadamard operations at the end of each round which turn the carriers into each other, Fig (\ref{nsf}).
 At the end of this round the carrier will be in the state (\ref{oddd}) and after the Hadamard operation again turns into
(\ref{new-rho-0}) or equivalently (\ref{rho-0}) and will be ready for the next round. \\

Since this protocol is linear, it can be used for quantum state sharing. A quantum state like $|\phi\ra=a|0\ra+b|1\ra$ can be encoded as $|\tilde{\phi}\ra_{1,2}=a|0,0\ra+b|1,1\ra$  in even rounds and as $|\tilde{\phi}\ra_{1,2}=a|{\bf 0}\ra+b|{\bf 1}\ra$ in odd rounds and communicated to the receivers who retrieve it in the original form with some error. As mentioned before, only odd rounds are used for sharing states, and the retrieved state is given by  $\rho_\phi=(1-p)|\phi\ra\la \phi|+pX|\phi\ra\la \phi|X$. Therefore in view of the definition of the complete channel in definition (\ref{defff}), we have now proved the following:\\

\noindent {\bf The overall effect of de-phasing noise on the complete channel:}  If the carrier is spoiled by noise as in (\ref{rho-0}), the complete quantum channel between the players $\Phi$ changes to

\be\label{comp1}
\Phi(\rho)==
\left\{\begin{array}{lr}
	\rho\h & {\rm even \ \ rounds}\\
	(1-p)\rho + \sigma_x\rho \sigma_x & {\rm odd \ \ rounds}
\end{array}\right.
\ee

\iffalse

To find the average fidelity, we use the Bloch representation of the state $|\phi\ra $ and the relations  $\la \phi|X|\phi\ra\equiv tr(\frac{1}{2}(1+{\bf n}\cdot \bm\sigma)\sigma_x)=n_x$ and $\int d{\bf n}n_x^2=\frac{1}{3}$ to arrive at
\be\label{fid}
F_{av}= \frac{1}{4\pi}\int \la \phi|\rho_
\phi|\phi\ra=1-p+\frac{1}{4\pi}\int \la \phi|X|\phi\ra^2=1-\frac{2p}{3}.
\ee
\fi

\section{Effect of depolarizing noise}\label{n3}
We now consider another important type of noise, namely a global depolarizing noise which affects the carrier as follows:
\be\label{ggg}
\rho^{(even)}_{_{A,B,C}}=(1-p)|G_0\ra\la G_0	|+\frac{p}{8}I,
\ee
where for simplicity of notation, we have denoted the standard GHZ state, simply as $G_0.$ The full set of  GHZ states which form a basis for the space of three qubits are:
\ba\label{Gs}
|G_0\ra_{_{A,B,C}}=\frac{1}{\sqrt{2}}(|000\ra+|111\ra),&&\h |G'_0\ra_{_{A,B,C}}=\frac{1}{\sqrt{2}}(|000\ra-|111\ra),\cr
|G_A\ra_{_{A,B,C}}=\frac{1}{\sqrt{2}}(|100\ra+|011\ra),&&\h |G'_A\ra_{_{A,B,C}}=\frac{1}{\sqrt{2}}(|100\ra-|011\ra),\cr
|G_B\ra_{_{A,B,C}}=\frac{1}{\sqrt{2}}(|010\ra+|101\ra),&&\h |G'_B\ra_{_{A,B,C}}=\frac{1}{\sqrt{2}}(|010\ra-|101\ra),\cr
|G_C\ra_{_{A,B,C}}=\frac{1}{\sqrt{2}}(|001\ra+|110\ra),&&\h |G'_C\ra_{_{A,B,C}}=\frac{1}{\sqrt{2}}(|001\ra-|110\ra).
\ea
These states form an ortho-normal basis, and have the following relations
 with each other
\be\label{GiX}
|G_i\ra_{_{A,B,C}}=X_i|G_0\ra_{_{A,B,C}},\h  |G'_i\ra_{_{A,B,C}}=X_i|G'_0\ra_{_{A,B,C}},\h
\ee
where $i=0,A,B,C$ and $X_0=I$. And they are complete
\be
I_{_{A,B,C}}=\sum_{i=0,A,B,C}|G_i\ra_{_{A,B,C}}\la G_i|+|G'_i\ra_{_{A,B,C}}\la G'_i|.
\ee
In order to see how the protocol runs in this case, we have to decompose the carrier in terms of the above collection of GHZ states
\be\label{Gnoisy}
\rho^{(even)}_{_{A,B,C}}=(1-p)|G_0\ra_{_{A,B,C}}\la G_0|+\frac{p}{8}\big[\sum_{i=0,A,B,C}|G_i\ra_{_{A,B,C}}\la G_i|+|G'_i\ra_{_{A,B,C}}\la G'_i|\big].
\ee

Using Eq. (\ref{GiX}), the relations $H^{\otimes 3}|G_0\ra=|{\bf 0}\ra$ and  $ZH=HX$, we also find
\be
H^{\otimes 3}|G_i\ra_{_{A,B,C}}=Z_i|{\bf 0}\ra_{_{A,B,C}},\h H^{\otimes 3}|G'_i\ra_{_{A,B,C}}=Z_i|{\bf 1}\ra_{_{A,B,C}},
\ee
where $Z_0=I$ and the above relations are correct modulo a global $\pm$ sign. \\

\begin{figure}[!ht]
	\centering
	\includegraphics[width=12.0cm,height=10cm,angle=0]{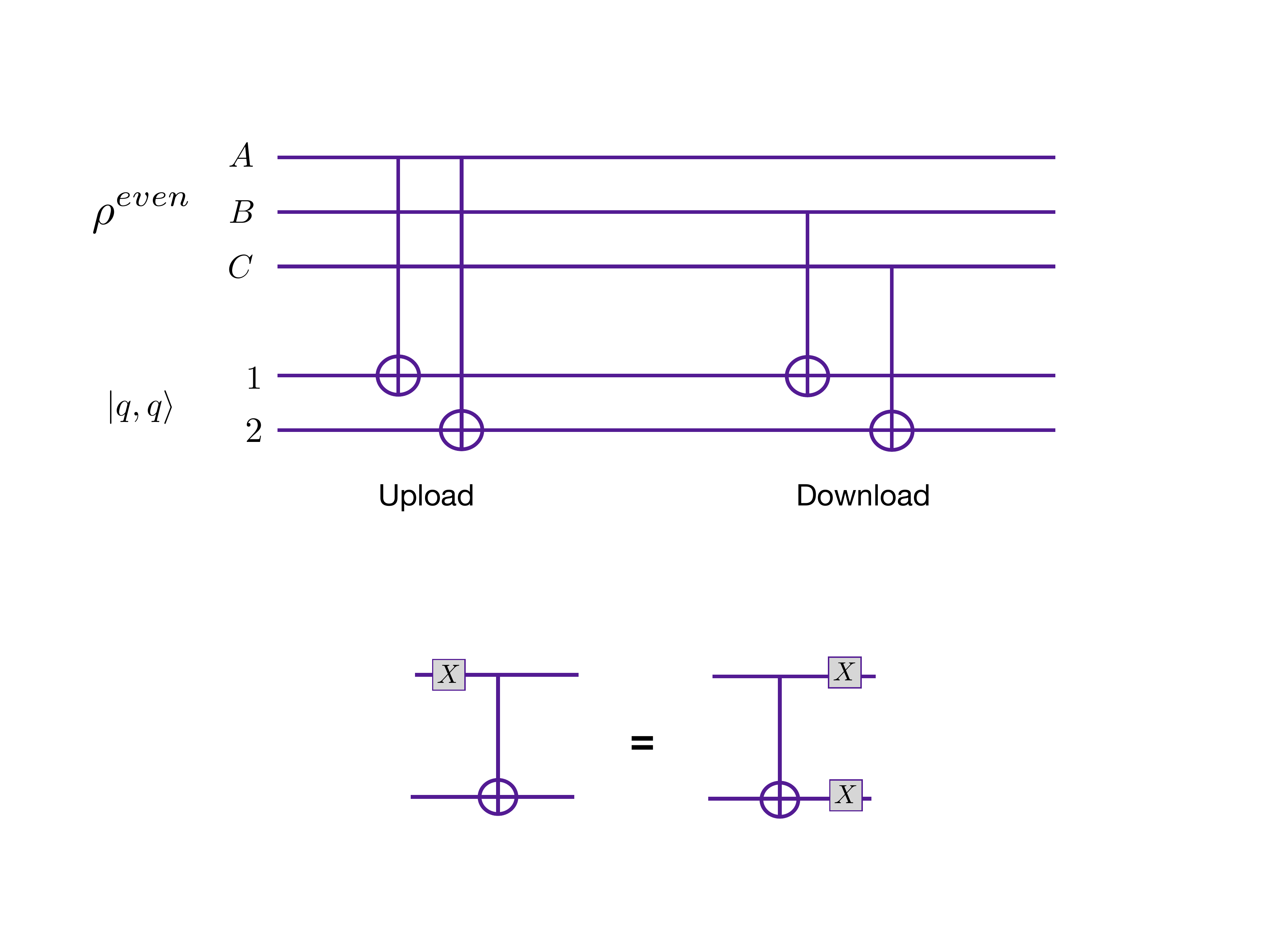}
	\caption{ The operations of players in even rounds shown in circuit form. The top figure shows encoding of the state $|q\ra$ and the actions of A, B and C and the bottom figure shows the identity ($C_{i,j}X_i=X_iX_jC_{i,j}$) which is used in deriving (\ref{Oeven}).
	}\label{CircuitEven}
\end{figure}

Equipped with these relations we now investigate how the protocol runs if the carrier is not a pure GHZ states but a mixture of different GHZ states in Eq. (\ref{Gs}). As performed for the de-phasing noise, we consider one of the GHZ states, run the protocol with this carrier and combine the results at the end. Let the carrier be $|G_i\ra$. The state to be shared is $|q,q\ra$, and the full actions of the players is given by
$\Omega^{even}=C_{_{C,2}}C_{_{B,1}}C_{_{A,2}}C_{_{A,1}}$, shown in the circuit diagram (\ref{CircuitEven}). Using the relation $C_{i,j}X_i=X_iX_jC_{i,j}$
shown in Fig. (\ref{CircuitEven}), we find
\ba\label{Oeven}
\Omega^{(even)}X_{_A}&=&X_{_A}X_{_1}X_{2}\Omega^{(even)},\cr  \Omega^{(even)}X_{_B}&=&X_{_B}X_{_1}\Omega^{(even)},\cr  \Omega^{(even)}X_{_C}&=&X_{_C}X_{_2}\Omega^{(even)}.
\ea
 \begin{figure}[!ht]
	\centering	\includegraphics[width=12.0cm,height=10cm,angle=0]{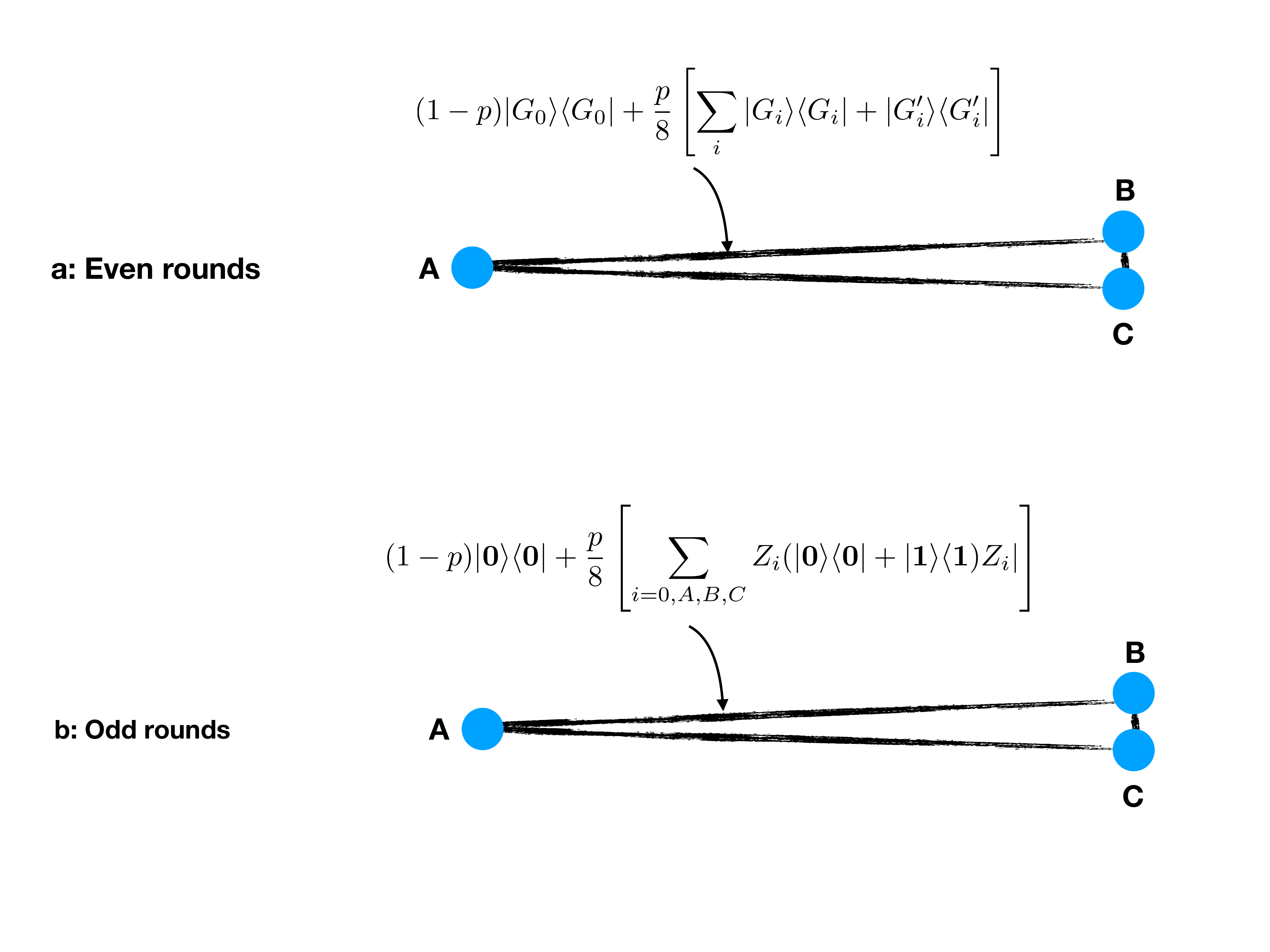}
	\caption{  Under a global depolarizing noise, the carrier alternates between the above two forms in even and odd rounds.
	}\label{nsf2}
\end{figure}

Running the protocol on $|G_i\ra_{_{A,B,C}}|q,q\ra_{1,2}$, using (\ref{Oeven}) and the above relations, we find

\be
\Omega^{(even)}_{_{A,B,C}}|G_i\ra_{_{A,B,C}}|q,q\ra_{_{1,2}}=|G_i\ra_{_{A,B,C}}X^{(i)}|q,q\ra_{_{1,2}},
\ee
where
\be
X^{(0)}=I\ \  ,  \ \ X^{(A)}=X_1X_2 \ \ , \ \  X^{(B)}=X_1 \ \ , \ \ X^{(C)}=X_2.
\ee
It is obvious that exactly the same relations hold when the carrier $|G_i\ra$ is replaced with $|G'_i\ra$.
 Combining all the relations we find that when the protocol is run on the noisy carrier (\ref{Gnoisy}), the result is:

\ba
\rho^{(even)}_{_{A,B,C}}\otimes |q,q\ra_{_{1,2}}\la q,q| &\lo& (1-p)|G_0\ra\la G_0|\otimes |q,q\ra\la q,q|\cr
&+&\frac{p}{8}\sum_{i}\bigg[\big(|G_i\ra\la G_i|+|G'_i\ra\la G'_i|\big)\otimes X^{(i)}|q,q\ra\la q,q|X^{(i)}\bigg].
\ea

Therefore with probability $P={1-p}+\frac{p}{4}=1-\frac{3p}{4}$, the state $|q,q\ra$ is transmitted in correct form. With probability $\frac{3p}{4}$ it is transmitted in a form where one or both of the qubits have been flipped. Thus the error probability for even rounds is equal to $P_{error}^{(even)}=\frac{3p}{4}$. \\

Consider now the odd rounds. The carrier is now given by
\be\label{rodd}
\rho^{(odd)}_{_{A,B,C}}=(1-p)|{\bf 0}\ra\la{\bf 0} |+\frac{p}{8}\bigg[\sum_{i=0,A,B,C}Z_i\big(|{\bf 0}\ra\la {\bf 0}|+|{\bf 1}\ra\la {\bf 1}|\big)Z_i\bigg],
\ee
obtained by the Hadamard actions of the players on (\ref{Gnoisy}),  Fig (\ref{nsf2}). Note that since $H^2=I$ and $HZ=XH$, the carriers in the even and odd rounds turn into each other by the Hadamard action of the players. \\

\begin{figure}[!ht]
	\centering
	\includegraphics[width=12.0cm,height=10cm,angle=0]{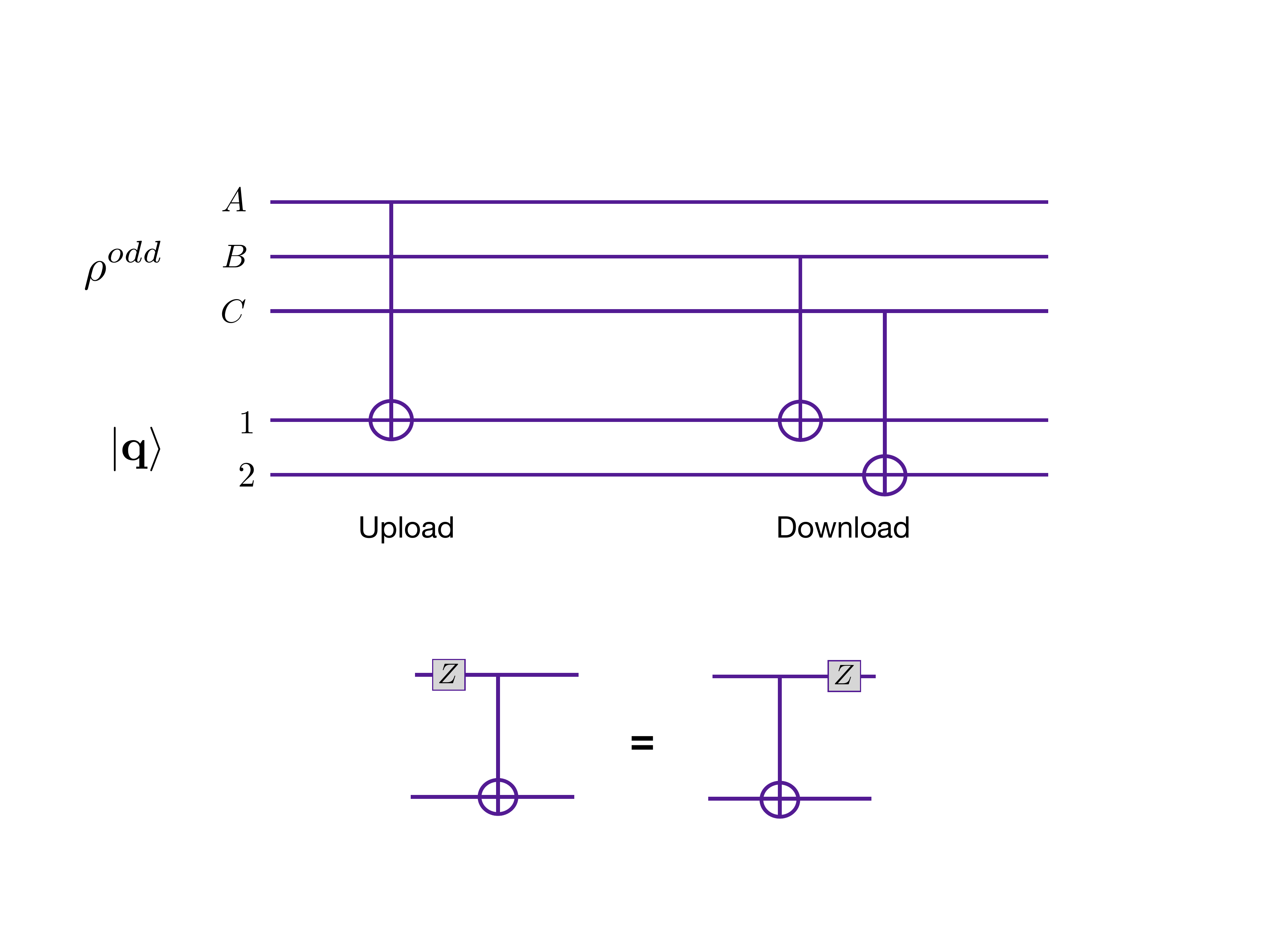}
	\caption{ The operations of players in odd rounds shown in circuit form. The top figure shows encoding of the state $|q\ra$ and the actions of A, B and C and the bottom figure shows the identity ($C_{i,j}Z_i=Z_iC_{i,j}$) which is used in deriving (\ref{OmegaOdd}).
	}\label{CircuitOdd}
\end{figure}

We now have to consider the carriers $Z_i|{\bf 0}\ra$ and $Z_i|{\bf 1}\ra$ separately and then combine the results. In odd rounds, the operations of the players is given by $\Omega^{odd}=C_{_{C,2}}C_{_{B,1}}C_{_{A,1}}$. This operation is shown in circuit diagram (\ref{CircuitOdd}) where we use the relation $Z_iC_{i,j}=C_{i,j}Z_i$, Fig.(\ref{CircuitOdd}), to find

\be\label{OmegaOdd}
\Omega^{odd}Z_i=Z_i \Omega^{odd},
\ee
from which we find

\ba
\Omega^{(odd)}Z_i|{\bf 0}\ra_{_{ABC}}|{\bf q}\ra_{_{1,2}}&=&Z_i|{\bf 0}\ra_{_{ABC}}|{\bf q}\ra_{_{1,2}},\cr
\Omega^{(odd)}Z_i|{\bf 1}\ra_{_{ABC}}|{\bf q}\ra_{_{1,2}}&=&Z_i|{\bf 1}\ra_{_{ABC}}|\overline{{\bf q}}\ra_{_{1,2}}.
\ea

Combining these equations and using Eq. (\ref{rodd})  we find

\ba
\rho^{(odd)}_{_{A,B,C}}\otimes |{\bf q}\ra_{1,2}\la {\bf q}| &\lo& (1-p)|{\bf 0}\ra\la {\bf 0}|\otimes |{\bf q}\ra\la {\bf q}|+
+\frac{p}{8}\sum_{i}\bigg[\big(Z_i|{\bf 0}\ra\la {\bf 0}|Z_i)\otimes |{\bf q}\ra\la {\bf q}|\bigg]\cr
&+&\frac{p}{8}\sum_{i}\bigg[\big(Z_i|{\bf 1}\ra\la {\bf 1}|Z_i)\otimes |\overline{{\bf q}}\ra\la \overline{{\bf q}}|\bigg].
\ea

This means that in odd rounds, the states $|{\bf q}\ra$ is transmitted correctly with probability $P=1-p+\frac{p}{2}=1-\frac{p}{2}$ and in flipped form $|\overline{{\bf q}}\ra$ with probability $P_{error}^{odd}=\frac{p}{2}$. Therefore we have shown the following:\\

\noindent {\bf The overall effect of de-polarizing noise on the complete channel:}  If the carrier is spoiled by depolarizing noise as in (\ref{ggg}), the complete quantum channel between the players $\Phi$ changes to

\be\label{comp2}
\Phi(\rho)==
\left\{\begin{array}{lr}
	(1-\frac{3p}{4})\rho +\frac{3p}{4}\sigma_x\rho\sigma_x\h & {\rm even \ \ rounds}\\
	(1-\frac{p}{2})\rho +\frac{p}{2} \sigma_x\rho \sigma_x & {\rm odd \ \ rounds}.
\end{array}\right.
\ee

\section{Further considerations}
In this section we discuss some related problems the first two of them are straightforward and the last one requires further investigation. 
\subsection{The problem of error correction}
What we have shown is that the effect of noisy carriers, when they are affected by de-phased or de-polarized into mixed states is quite simple in the sense that the complete channel of definition (\ref{defff}) becomes a bit-flip channel with different parameters in even and odd rounds as in (\ref{comp1}) and (\ref{comp2}). Therefore many of the techniques used for quantum error correction, i.e. encoding the states into a larger Hilbert space and using syndrome measurement, can be done to achieve reliable quantum communication. \\

\subsection{Generalization to n-parties}
The original noiseless protocol  has been generalized to $(1,n)$ schemes where Alice shares her state with $n$ parties  \cite{marvian}.   Although the reader can find all the details of this generalization in \cite{marvian}, here we present a few basic facts which help the reader to reconstruct the whole scheme. First we note that the basic property of alternating carriers between $|GHZ_n\ra$ state and the even state $|{\bf 0}_n\ra$ states by the action of joint Hadamard actions remains intact, i.e.

\be
H^{\otimes n}|GHZ_n\ra=|{\bf 0}_n\ra\h H^{\otimes n}|{\bf 0}_n\ra=|GHZ_n\ra, 
\ee
where $|{\bf 0}_n\ra$ is  defined by induction as in (\ref{33}). Furthermore, 
any state $|\psi\ra=\a|0\ra+b|1\a$ is encoded into $\a|0^{\otimes n}\ra+b|1^{\otimes n}\ra$ in odd rounds (when the carrier is $|GHZ_n\ra$) and into $a|{\bf 0}_n\ra+b|{\bf 1}_n\ra$ in even rounds (when the carrier is $|{\bf 0}_n\ra$ ). \\

\subsection{The problem of security} Unlike quantum key distribution where there are information theoretic proofs of security of the schemes, for quantum secret and state sharing, no general and widely accepted proofs are available \cite{adesso}. Therefore no scheme can claim to be absolutely secure and any scheme of this sort is prone to attacks and cheating among participants. This is true for the multitude of schemes that are have been reported so far \cite{ZhangLiMan,chin1, chin2, chin3,chin4}. As for the present scheme, some of these attacks, like an outsider Eve entangling herself with the carrier has been discussed in the original proposal \cite{BK} and the following  literature \cite{marvian, attack1, attack2}, where improvements have also been proposed to prevent these attacks.  Here we have focused on one important aspect of this scheme to see how the whole communication is affected if the main carriers are no longer pure states and have been contaminated by two common types of noise.

\section{Conclusion}
The main purpose  of this paper is to study the theoretical possibility of using entangled states as reusable carriers of information. Therefore our concern has not been practical or security issues of protocols based on this idea. Rather we have studied the robustness of quantum state sharing protocol with reusable entanglement against de-phasing and depolarizing noise. In this type of protocol, the carrier which is a shared entangled state between the legitimate parties acts as a medium to which message states are entangled and disentangled in the sender and receiver ends. During the transition the state identity, being in a highly mixed state is hidden from adversaries. For security of the protocol, the carrier should alternate between two different types and it is crucial that these carriers are not distorted and destroyed completely by noise and the repeated downloading and uploading operations of the parties. The overall effect is that the channel transmits the basis states with a fixed and known error probability that we have calculated. One can then use quantum error correcting codes to transmit quantum states through this channel with as high fidelity as desired at the expense of lowering the transmission rate.  Finally we should mention that we have proved the robustness against the two most common types of noise, namely de-phasing and de-polarizing noises. It remains to be seen to what extent this scheme is robust against a general form of noise.  \\

\section{Acknowledgements}
This research was partially supported by a grant no. 96011347 from the Iran National Science Foundation. The work of V. K. was also partially supported by the  grant G950222 from the research grant system of Sharif Univeristy of Technology. We also thank Abdus Salam ICTP where the final stages of this work was completed.

{}

\end{document}